\documentclass[10pt,final,letterpaper,twocolumn,romanappendices]{IEEEtran}
\usepackage{cite}
\usepackage{epsfig}
\usepackage{amsmath,amssymb,amsfonts,amstext,amsbsy,amsopn,eucal,dsfont}
\usepackage{psfrag}
\usepackage{subfigure}
\usepackage{array}

\newtheorem{theorem}{\textbf{Theorem}}

\newcommand{\defn}{\triangleq}
\newcommand{\dif}{\textmd{d}}

\begin{document}

\title{\huge The Mean SIR of Large-Scale Wireless Networks: Its Closed-Form Expression and Main Applications}

\author{Chun-Hung Liu
\thanks{C.-H. Liu is with the Department of Electrical and Computer Engineering at National Chiao Tung University, Hsinchu, Taiwan. The contact author is Dr. Liu  (e-mail: chungliu@nctu.edu.tw). Manuscript Date: \today.}
}
\maketitle

\begin{abstract}
In a large-scale wireless ad hoc network in which all transmitters form a homogeneous of Poisson point process, the statistics of the signal-to-interference ratio (SIR) in prior work is only derived in closed-form for the case of Rayleigh fading channels. In this letter, the mean SIR is found in closed-form for general random channel (power) gain, transmission distance and power control models. According to the derived mean SIR, we first show that channel gain randomness actually benefits the mean SIR so that the upper bound on the mean spectrum efficiency increases. Then we show that stochastic power control and opportunistic scheduling that capture the randomness of channel gain and transmission distance can significantly not only enhance the mean SIR but reduce the outage probability. The mean-SIR-based throughput capacity is proposed and it can be maximized by a unique optimal intensity of transmitters if the derived supporting set of the intensity exists. 
\end{abstract}


\section{Introduction}
The statistics of the signal-to-interference ratio (SIR) of an interference-limited wireless network is very useful to help us comprehend how the transmission performance metrics, such as outage probability and throughput, vary with different random models of channel (power) gain, power control and spacial distribution of interfering nodes. However, it is hardly able to be theoretically acquired in closed-form especially when the wireless network inherently has general and complex spacial randomness. For example, an ad hoc network in which all transmitters form a homogeneous Poisson point process (PPP) is of general spacial randomness whereas its SIR statistics can be found in closed-form only for the case of Rayleigh fading channels \cite{FBBBPM06,MHRKG09,SWJGANJ07}. 

The difficulty of finding the statistics of the SIR in closed-form for a Poisson wireless network without Rayleigh fading is due to the the derivation processes without involving the Laplace transform of the interference. Unfortunately, this derivation difficulty seems not possible to be casted off until now. As a result, we resort to finding the mean SIR since it can also characterize a certain level of the randomness impact on the transmitting performance of a transmitter. Prior works in \cite{SWJGANJ07,CHLJGA11} on the statistics of the SIR in Poisson wireless networks with general channel fading merely characterize the bounds on the distribution of the SIR. There are no prior works on the mean SIR of a Poisson wireless network with general random channel gain models. 
 
In this letter, the mean SIR of a Poisson ad hoc network with general statistical models of channel gain, power control and transmission distance is derived in closed-form by applying a special integration technique on the Laplace transform of the interference. To the best of our knowledge, it is the first and very general work on the mean SIR in the literature.  According to the expression of the derived mean SIR, we first show that channel gain randomness actually benefits the mean SIR and  the outage probability and mean SIR both can be significantly improved if power control and/or opportunistic scheduling schemes are well-designed to catch up channel gain randomness.  The throughput capacity based on the mean SIR is proposed and it is essentially the upper bound on the mean Shannon-sense spectrum efficiency if interference is treated as noise. Finally, we show that the optimal transmitter intensity that maximizes the throughput capacity uniquely exists if its derived supporting set is nonempty.  

\section{System Model}
Consider a large-scale and interference-limited wireless ad hoc network in which all transmitters form a marked homogeneous PPP $\Phi$ of intensity $\lambda$ given by
\begin{equation}
\hspace{-1.2mm}\Phi \defn \{(X_i, D_i, P_i, H_i) :X_i\in\mathbb{R}^2, D_i\geq 1, P_i, H_i\in\mathbb{R}_{+}\},
\end{equation}
where $P_i$ denotes the (random) transmit power of transmitter $X_i$, $H_i$ characterizes the channel gain (such as fading and/or shadowing) between transmitter $X_i$ and the origin, and $D_i$ stands for the (random) transmission distance between transmitter $X_i$ and its desired receiver.  All $H_i$'s are independent and identically distributed (i.i.d.) random variables with unit mean, $P_i$'s are i.i.d. and $D_i$'s are i.i.d. as well if they are random. Consider a reference receiver located at the origin and its signal-to-interference ratio (SIR) $\gamma_0$  is given by
\begin{equation}\label{Eqn:SINR}
\gamma_0=\frac{P_0H_0 D_0^{-\alpha}}{I_0},
\end{equation}  
where $I_0=\sum_{X_i\in\Phi\setminus X_0} P_i G_i \|X_i\|^{-\alpha}$ represents the interference power at the reference receiver where $X_0$ is the transmitter of the reference receiver, $\alpha>2$ is the path loss exponent, $\|X-Y\|$ denotes the Euclidean distance between nodes $X$ and $Y$, and $G_i$ characterizes the channel gain from interferer $X_i$ to the origin. All $G_i$'s are i.i.d. as well.

According to the Slivnyak theorem \cite{DSWKJM96}, the interference powers evaluated at any nodes in the network have the same statistics if the nodes are a homogeneous PPP. That indicates the statistics of the SIR at any receiver in the network is also identical since all received signal powers are iid as well. Intuitively, the distribution of the point process is unaffected by the addition of a receiver at the origin. Accordingly, without loss of generality we will perform our analysis based on the reference receiver at the origin. The performance measured at the origin is often referred to the Palm measure, and to keep with simplified notation we will denote the probability and expectation of functionals evaluated at the origin by $\mathbb{P}$ and $\mathbb{E}$, respectively.

\section{The Closed-Form Mean SIR and Upper Bound on Area Spectrum Efficiency} 
In this section, the exact closed-form of the mean SIR in \eqref{Eqn:SINR} is first characterized for any random channel gain, power control and transmission distance models and afterward its important applications in power control, scheduling and upper bound on the mean spectrum efficiency are discussed. The mean SIR of any receivers in the network with general random models is shown in the following theorem.
\begin{theorem}\label{Thm:MeanSIR}
Under arbitrary channel gain, transmission distance and power control models, the mean SIR of any receivers in the network can be shown as
\begin{equation}\label{Eqn:AvgSIR}
\mathbb{E}[\gamma_0] = \frac{\mathbb{E}[PHD^{-\alpha}]\Gamma(1+\frac{\alpha}{2})}{\{\pi\lambda \Gamma\left(1-\frac{2}{\alpha}\right)\mathbb{E}[P^{\frac{2}{\alpha}}]\mathbb{E}[G^{\frac{2}{\alpha}}]\}^{\frac{\alpha}{2}}},
\end{equation}
where $\Gamma(x)=\int_{0}^{\infty} t^{x-1}e^{-t}\dif t$ is the gamma function.
\end{theorem}
\begin{IEEEproof}
Since the signal and interference powers in \eqref{Eqn:SINR} are independent, the mean of $\gamma_0$ can be written as
\begin{align}
\mathbb{E}[\gamma_0] = \mathbb{E}\left[HPD^{-\alpha}\right]\cdot \mathbb{E}\left[\frac{1}{I_0}\right].\label{Eqn:ProofAvgSIR}
\end{align}
The mean of $1/I_0$ can be found as follows
\begin{align*}
\mathbb{E}\left[\frac{1}{I_0}\right]&=\mathbb{E}\left[\int_{0}^{\infty} e^{-sI_0}\dif s\right] = \int_{0}^{\infty} \mathbb{E}\left[e^{-sI_0}\right]\dif s.
\end{align*}
According to the probability generating functional (PGF) of a homogeneous PPP \cite{MHRKG09,DSWKJM96}, letting $Y$ be an exponential random variable with unit mean leads to
\begin{align*}
\mathbb{E}\left[e^{-sI_0}\right]&=\mathbb{E}\left[e^{-s\sum_{X_i\in\Phi\setminus X_0}P_iG_i\|X_i\|^{-\alpha}}\right]\\
&=\mathbb{E}_{\Phi}\left\{\mathbb\prod_{X_i\in\Phi}\mathbb{E}_{P_iG_i}\left[e^{-sP_iG_i\|X_i\|^{-\alpha}}\right]\right\}\\
&\stackrel{(\star)}{=} \exp\left(-\pi\lambda\mathbb{E}\left[ \int_{0}^{\infty}\left(1-e^{-sPG r^{-\frac{\alpha}{2}}}\right) \dif r\right]\right) \\
&=\exp\left(-\pi\lambda \int_{0}^{\infty}\mathbb{P}[Y\leq sPGr^{-\frac{\alpha}{2}}] \dif r\right) \\
&=\exp\left(-\pi\lambda\Gamma\left(1-\frac{2}{\alpha}\right)s^{\frac{2}{\alpha}} \mathbb{E}[P^{\frac{2}{\alpha}}]\mathbb{E}[G^{\frac{2}{\alpha}}]\right),
\end{align*}
where $(\star)$ is due to the PGF of a homogeneous PPP. Thus,
\begin{align*}
\mathbb{E}\left[\frac{1}{I_0}\right]&=\int_{0}^{\infty} \exp\left(-\pi\lambda\Gamma\left(1-\frac{2}{\alpha}\right)s^{\frac{2}{\alpha}} \mathbb{E}[P^{\frac{2}{\alpha}}]\mathbb{E}[G^{\frac{2}{\alpha}}]\right)\dif s\\ &=\frac{\Gamma(1+\frac{\alpha}{2})}{\{\pi\lambda \Gamma\left(1-\frac{2}{\alpha}\right)\mathbb{E}[P^{\frac{2}{\alpha}}]\mathbb{E}[G^{\frac{2}{\alpha}}]\}^{\frac{\alpha}{2}}}.
\end{align*}
Substituting the above result into \eqref{Eqn:ProofAvgSIR} results in \eqref{Eqn:AvgSIR}. 
\end{IEEEproof} 

\begin{figure}[!t]
\centering
\includegraphics[width=3.5in, height=2.25in]{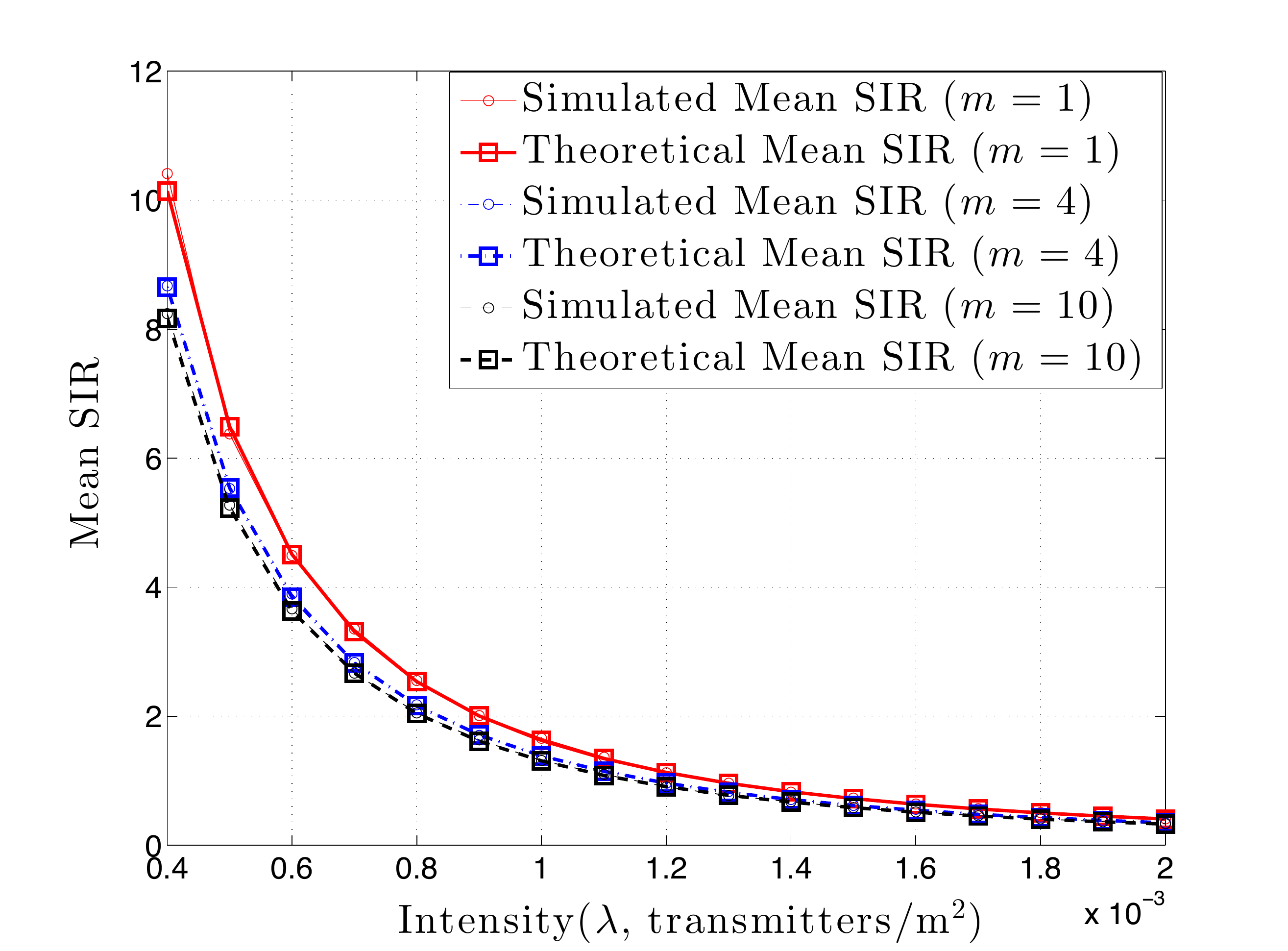}
\caption{The simulation and theoretical results of the mean SIR for $\alpha=4$ and $D=15$m. The mean SIR slightly decreases when $m$ increases (i.e. channels become less fading.), i.e. fading indeed benefits the mean SIR.}
\label{Fig:MeanSIRFading}
\end{figure}

The closed-form mean of the SIR in Theorem \ref{Thm:MeanSIR} is a very general result which is different from the existing ones obtained by assuming Rayleigh fading channels \cite{MHRKG09}. Most importantly, it is valid for any stochastic power control, channel power, and transmission distance models. For example, if all transmit powers are the same constant, all transmission distances are constant $d$, and all channels undergo Nakagami fading and their fading gains are i.i.d. Gamma distribution with unit mean and variance $1/m$ for any $m>0$, then we have $\mathbb{E}[H]=1$, $\mathbb{E}[G^{\frac{2}{\alpha}}]=\Gamma(m+\frac{2}{\alpha})/\Gamma(m)m^{\frac{2}{\alpha}}$ and the mean SIR in \eqref{Eqn:AvgSIR} can be explicitly expressed as
\begin{align}
\mathbb{E}[\gamma_0(m)]=\frac{m\Gamma(1+\frac{\alpha}{2})[\Gamma(m)]^{\frac{\alpha}{2}}}{[\pi d^2\lambda\Gamma(1-\frac{2}{\alpha})\Gamma(m+\frac{2}{\alpha})]^{\frac{\alpha}{2}}},\,m>0.
\end{align}
Since we can show that $\mathbb{E}[\gamma_0(m)]$ is a monotonic decreasing and convex function of $m$, we have $\mathbb{E}[\gamma_0(m)]\geq \mathbb{E}[\gamma_0(\infty)]$, which indicates that \textit{fading benefits the mean SIR since $\mathbb{E}[\gamma_0(\infty)]$ corresponds to the mean SIR without fading}. The intuition behind this observation is because fading only weakens the interference channel powers. The simulation results of the mean SIR in Fig. \ref{Fig:MeanSIRFading} apparently support this  observation.  

The mean SIR can be used to acquire the exact upper bound on the mean spectrum efficiency that in general cannot be found in closed-form, especially in a Poisson wireless network.  The following theorem specifies the exact upper bound on the mean Shannon channel capacity per unit spectrum and when the bound is tight.
\begin{theorem}\label{Thm:UppBoundMeanRate}
If the interference power is treated as noise, the upper bound on the mean spectrum efficiency $C(\lambda)\defn\mathbb{E}[\log_2(1+\gamma_0)]$ is given by
\begin{align}\label{Eqn:UppBoundLinkCapa}
C(\lambda)\leq \log_2(1+\mathbb{E}[\gamma_0])=\log_2\left(1+\frac{\kappa}{\lambda^{\frac{\alpha}{2}}}\right),
\end{align}
where $\kappa\defn \frac{\mathbb{E}[PHD^{-\alpha}]\Gamma(1+\frac{\alpha}{2})}{\{\pi \Gamma\left(1-\frac{2}{\alpha}\right)\mathbb{E}[P^{\frac{2}{\alpha}}]\mathbb{E}[G^{\frac{2}{\alpha}}]\}^{\frac{\alpha}{2}}}$ and the upper bound is fairly tight whenever $\lambda\gg \kappa^{\frac{2}{\alpha}}$.
\end{theorem}
\begin{IEEEproof}
By Jensen's inequality, the upper bound on $C(\lambda)$ is $\log_2(1+\mathbb{E}[\gamma_0])$. Substituting \eqref{Eqn:AvgSIR} into $\log_2(1+\mathbb{E}[\gamma_0])$ leads to \eqref{Eqn:UppBoundLinkCapa}. To make the upper bound tight, $\mathbb{E}[\gamma_0]\ll 1$ should hold since $C(\lambda)=\mathbb{E}[\log_2(1+\gamma_0)]\approx \frac{\mathbb{E}[\gamma_0]}{\ln 2}$ for $\gamma_0\ll 1$ a.s. Thus, $\kappa \lambda^{-\frac{\alpha}{2}}\ll 1$, i.e., $\lambda\gg \kappa^{\frac{2}{\alpha}}$.
\end{IEEEproof}
According to Theorem \ref{Thm:UppBoundMeanRate}, we realize that $C(\lambda)$ can be accurately represented by its upper bound provided that $\kappa^{\frac{2}{\alpha}}$ is substantially smaller than the intensity. Namely, the situations, such as long transmission distance, large path loss and dense network, etc., are able to make the bound tight. There are more implications can be grasped from Theorems \ref{Thm:MeanSIR} and \ref{Thm:UppBoundMeanRate}, and they result in a couple of pivotal applications, which will be specified in the following section. 

\section{Main Applications of the Mean SIR}
The closed-form expression of the mean SIR in Theorem \ref{Thm:MeanSIR} reveals that power control and channel gain models have a significant impact on the magnitude of the mean SIR. We first study the fundamental interactions among mean SIR, power control, and opportunistic scheduling. Afterwards, we study the throughput capacity that is defined on the basis of the mean SIR.
 
\subsection{Power Control and Opportunistic Scheduling}
\begin{figure}[!t]
\centering
\includegraphics[width=3.5in, height=2.25in]{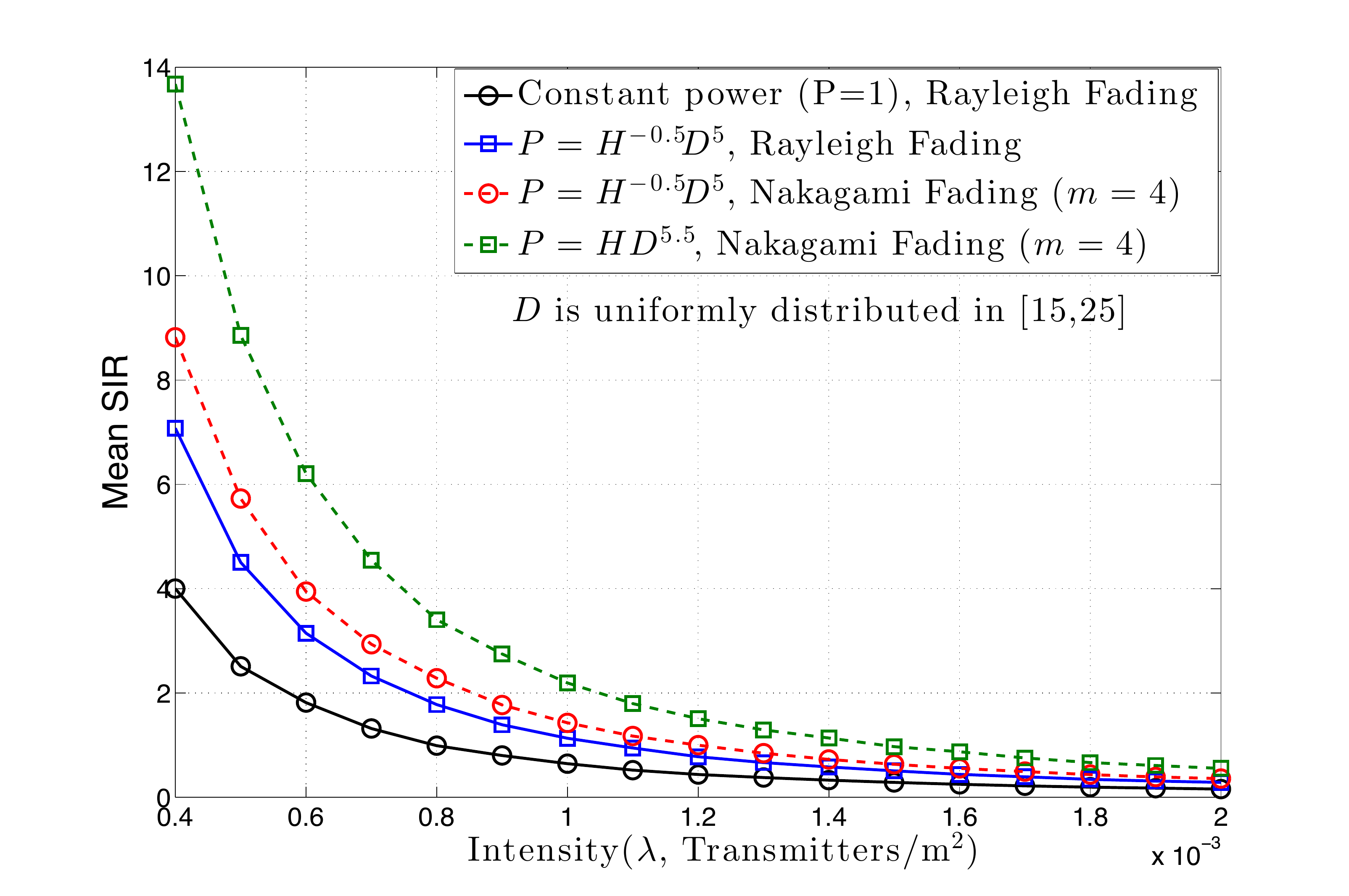}
\caption{The simulation results of the mean SIR with $P=H^{\rho}D^{\upsilon}/\mathbb{E}[H^{\rho}D^{\upsilon}]$ for $\alpha=4$, Nakagami-$m$ fading and uniformly distributed $D\in[15,25]$. The mean SIR increases as $P$ satisfies the conditions in Theorem \ref{Thm:PowerControl}.}
\label{Fig:MeanSIRPowerControl}
\end{figure}

\begin{figure}[!t]
	\centering
	\includegraphics[width=3.5in, height=2.25in]{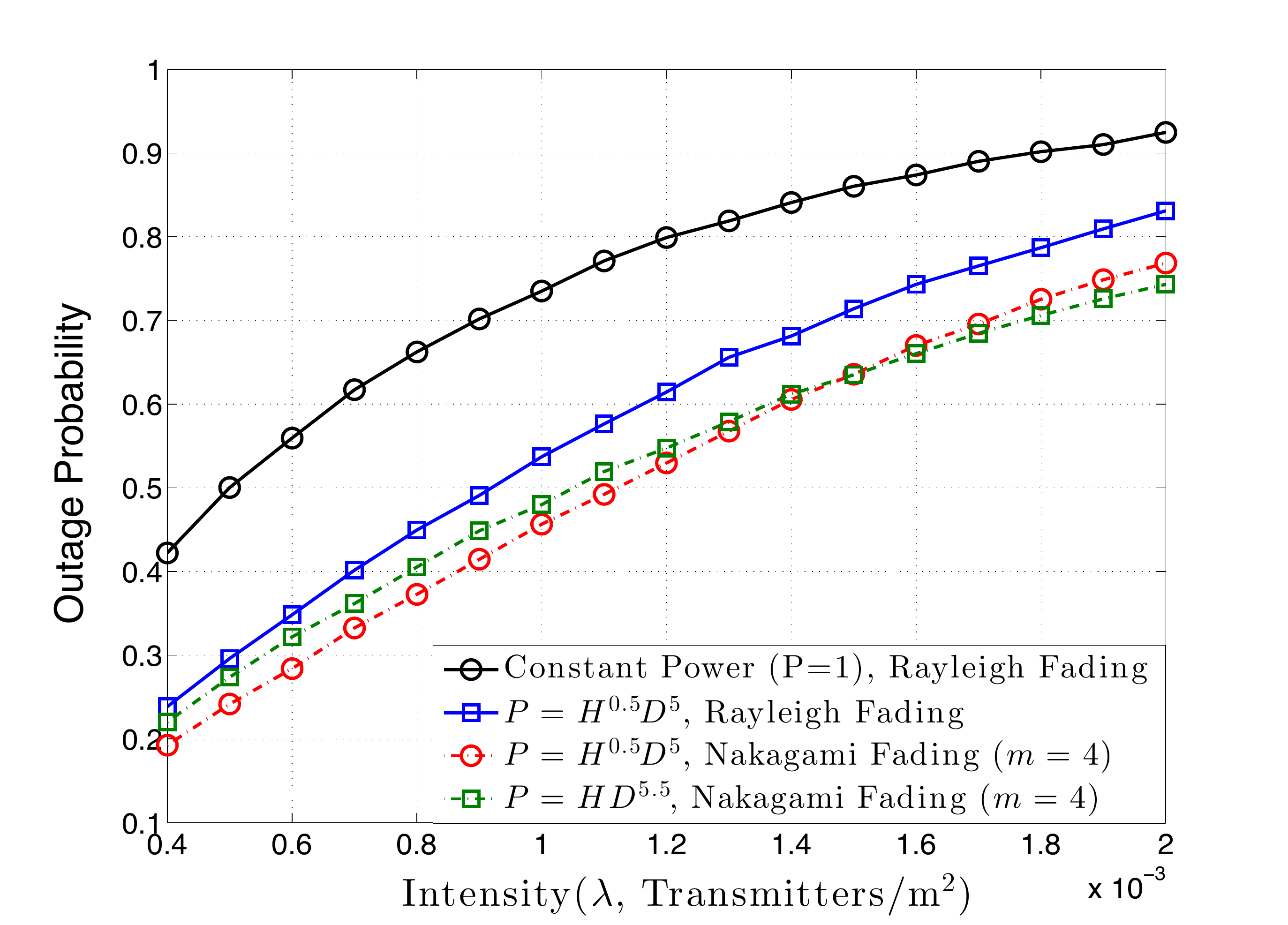}
	\caption{The simulation results of the outage probability with $P=H^{\rho}D^{\upsilon}/\mathbb{E}[H^{\rho}D^{\upsilon}]$ for $\alpha=4$, $\theta=0.5$, Nakagami-$m$ fading and uniformly distributed $D\in[15,25]$. The outage probability decreases as $P$ satisfies the conditions in Theorem \ref{Thm:PowerControl}.}
	\label{Fig:MeanSIROutageProb}
\end{figure}

Power control is one of the most important factors that deeply affect the mean SIR given in \eqref{Eqn:AvgSIR}. To elaborate this, first consider the case that any stochastic power control policies that do not depend on the channel gain and path loss. In this case, the mean SIR is virtually affected by the term $(\mathbb{E}[P])^{\frac{2}{\alpha}}/\mathbb{E}[P^{\frac{2}{\alpha}}]$ which is always greater than unity by H\"{o}lder's inequality. Accordingly, we learn that \textit{any non-channel-aware stochastic power control policies benefit the mean SIR}. Accordingly, we can conjecture that channel-aware power control also increases the mean SIR if it is appropriately designed based on the channel gains, which is validated in the following theorem.
\begin{theorem}\label{Thm:PowerControl}
Suppose all transmitters can obtain channel state information from their receivers and adopt the channel-aware stochastic power control law, $P\in \Theta(H^{\rho} D^{\upsilon})$ where $\rho$ and $\upsilon$ are both constants, with finite mean $\mathbb{E}[P]<\infty$. The power control law  increases the mean SIR of a receiver if $\rho\geq -1$, $\mathbb{E}[H^{1+\rho}]\geq\mathbb{E}[H^{\rho}]$, $\upsilon\geq \alpha$ and $\mathbb{E}[D^{1-\frac{\alpha}{\upsilon}}]\geq \mathbb{E}[D]$. 
\end{theorem}
\begin{IEEEproof}
According to the H\"{o}lder inequality for a positive random variable $Z$, we have $\mathbb{E}[Z^a]^{\frac{1}{c}}\geq \mathbb{E}[Z^{\frac{a}{c}}]$ if $a>0$ and $c>1$. If $\mathbb{E}[Z^{a}]\geq \mathbb{E}[Z^b]$ for $a\geq b\geq 0$, we further have $\mathbb{E}[Z^a]^{\frac{1}{c}}\geq \mathbb{E}[Z^{\frac{a}{c}}]\geq\mathbb{E}[Z^{\frac{b}{c}}]$. Now if the power control law $P\in\Theta(H^{\rho} D^{\upsilon})$ is applied, the mean SIR becomes
\begin{align}
\mathbb{E}[\gamma_0]=\frac{\mathbb{E}[H^{\rho+1}]\mathbb{E}[D^{\upsilon-\alpha}]\Gamma(1+\frac{\alpha}{2})}{\{\pi\lambda \Gamma\left(1-\frac{2}{\alpha}\right)\mathbb{E}[H^{\frac{2}{\alpha}\rho}]\mathbb{E}[D^{\frac{2}{\alpha}\upsilon}]\mathbb{E}[G^{\frac{2}{\alpha}}]\}^{\frac{\alpha}{2}}},
\end{align}
which increases because $\mathbb{E}[H^{\rho+1}]^{\frac{2}{\alpha}}\geq \mathbb{E}[H^{\frac{2}{\alpha}\rho}]$ due to $\rho\geq -1$ and $\mathbb{E}[H^{1+\rho}]\geq\mathbb{E}[H^{\rho}]$ whereas $\mathbb{E}[D^{\upsilon-\alpha}]^{\frac{2}{\alpha}}\geq \mathbb{E}[D^{\frac{2}{\alpha}\upsilon}]$ due to $\upsilon\geq\alpha$ and $\mathbb{E}[D^{\upsilon(1-\frac{\alpha}{\upsilon})}]\geq\mathbb{E}[D^{\upsilon}]$.
\end{IEEEproof}
Theorem 1 essentially expounds  that \textit{the mean SIR increases if power control can capture the randomness of channel and distance variations}. Accordingly, for a fixed SIR threshold $\theta$ the outage probability $\mathbb{P}[\gamma_0<\theta]$ can be reduced by the power control law since the mean SIR $\mathbb{E}[\gamma_0]=\int_{0}^{\infty} (1-\mathbb{P}[\gamma_0<\theta])\dif\theta$ is increased and outage probability is a monotonic increasing function of threshold $\theta$. This important observation contradicts the general consensus that power control in an interference-limited network is not an efficient means of enhancing transmission performance in terms of outage probability or SIR. In addition, the outage probability in a Poisson ad hoc network with channel-aware power control cannot be found in closed-form so that how it is precisely affected by power control cannot be analyzed. The simulation results in Figs. \ref{Fig:MeanSIRPowerControl} and \ref{Fig:MeanSIROutageProb} respectively present the mean SIR and outage probability for Nakagami-$m$ fading, uniform-distributed transmission distance and power $P=H^{\rho}D^{\upsilon}/\mathbb{E}[H^{\rho}D^{\upsilon}]$ and they obviously illustrate that channel-aware power control can significantly improve the mean SIR as well as outage probability.

The performance of power control on the mean SIR can be further improved by opportunistic scheduling. Consider every transmitter adopts a channel-aware opportunistic scheduling scheme that schedules a transmission whenever its channel gain is larger than some threshold \cite{SWJGANJ07,CHLYCT13}. For example, if a transmitter opportunistically schedules a transmission under the constraints $H\geq h_0$ and $D \leq d_0$ for two constants $h_0>0,d_0>1$ (i.e., it avoids transmitting in the situation of deep fading and long distance), the resulting transmitting nodes consist of a homogeneous PPP of intensity $\lambda \mathbb{P}[H\geq h_0]\mathbb{P}[D\leq d_0]$ since all $H_i$'s and $D_i$'s are independent. Under this scheduling scheme, the mean SIR with power control law $\mathbb{P}\in\Theta(H^{\rho}D^{\upsilon})$ can be derived as
\begin{align*}
\mathbb{E}[\gamma_0]=\frac{\mathsf{E}^c_H(\rho+1,h_0)\mathsf{E}_D(\upsilon-\alpha,d_0)\Gamma(1+\frac{\alpha}{2})}{\{\pi\lambda_s \Gamma\left(1-\frac{2}{\alpha}\right)\mathsf{E}^c_H(\frac{2}{\alpha}\rho,h_0)\mathsf{E}_D(\frac{2}{\alpha}\upsilon,d_0)\mathbb{E}[G^{\frac{2}{\alpha}}]\}^{\frac{\alpha}{2}}},
\end{align*}
where $\lambda_s=\lambda F_H^c(h_0)F_D(d_0)$, $\mathsf{E}^c_Z(a,b)\defn \mathbb{E}[Z^a|Z\geq b]=[bF^c_Z(b^{\frac{1}{a}})+\int_{b}^{\infty}F^c_Z(z^{\frac{1}{a}})\dif z]/F^c_Z(b)$, $\mathsf{E}_Z(a,b)\defn\mathbb{E}[Z^a|Z\leq b]=\int_{0}^{b} F^c_Z(z^{\frac{1}{a}})\dif z/F_Z(b)$ and $F_Z(\cdot) (F^c_Z(\cdot))$ is the CDF (CCDF) of random $Z$. Apparently, the opportunistic scheduling scheme not only increases the received signal power but also reduces the interference power as long as $F_H^c(h_0)F_D(d_0)\mathsf{E}^c_H(\frac{2}{\alpha}\rho,h_0)\mathsf{E}_D(\frac{2}{\alpha}\upsilon,d_0)<\mathsf{E}^c_H(\frac{2}{\alpha}\rho,0)\mathsf{E}_D(\frac{2}{\alpha}\upsilon,\sup(D))$. Thus, a pair of well-chosen threshold values of $h_0$ and $d_0$ is able to significantly improve the mean SIR as well as the outage probability.  

\subsection{Throughput Capacity and Its Optimality}
According to the upper bound on the mean spectrum efficiency given in Theorem \ref{Thm:UppBoundMeanRate}, we define the throughput capacity of a wireless network with an SIR threshold $\theta$ as follows
\begin{align}
\mathcal{T}(\lambda,\theta) \defn \log_2(1+\mathbb{E}[\gamma_0])\lambda\mathbb{P}[\gamma_0\geq \theta] \label{Eqn:ThrputCap}
\end{align}
whose physical meaning is the upper bound on the area spectrum efficiency $\lambda C(\lambda)F^c_{\gamma_0}(\lambda,\theta)$.  Although the throughput capacity is similar to the definition of spatial throughput or transmission capacity of a wireless network in \cite{FBBBPM06,SWXYJGAGDV05}, its feature is to capture  how the channel capacity changes with the intensity under different power control and opportunistic scheduling schemes, which is ignored in the previous network throughput metrics proposed in the literature. Consequently, the throughput capacity can characterize the fundamental limit of  the per-unit-area network throughput in a Shannon-capacity sense, and its optimality can be attained if a unique optimal intensity exists as shown in the following theorem. 
\begin{theorem}\label{Thm:OptThrputCap}
Let set $\Pi_{\lambda}$ define as
\begin{align}
\Pi_{\lambda}\defn\left\{\lambda\in\mathbb{R}_{++}: \frac{\lambda}{2}\frac{\dif^2\ell}{\dif\lambda^2}<-\frac{\dif\ell}{\dif\lambda}<\frac{\ell}{\lambda}\right\},
\end{align}
where $\ell(\lambda,\theta)=F^c_{\gamma_0}(\lambda,\theta)\log_2(1+\kappa\lambda^{-\frac{\alpha}{2}})$ for all $\lambda\in\mathbb{R}_{++}$. If $\Pi_{\lambda}\neq\emptyset$, there exists an unique optimal intensity $\overline{\lambda}_*\in\Pi_{\lambda}$ that maximizes the throughput capacity defined in \eqref{Eqn:ThrputCap}.
\end{theorem}
\begin{IEEEproof}
Since $\mathcal{T}(\lambda,\theta) =\lambda F^c_{\gamma_0}(\lambda,\theta)\log_2\left(1+\kappa \lambda^{-\frac{\alpha}{2}}\right)=\lambda\ell(\lambda,\theta)$ and $F^c_{\gamma_0}(\lambda,\theta)$ and $\log_2(1+\kappa\lambda^{-\frac{\alpha}{2}})$ are both monotonic decreasing and convex function of $\lambda$, their product $\ell(\lambda,\theta)$ is also monotonic and convex. 
The first and second derivatives of $\mathcal{T}(\lambda,\theta)$ with respect to $\lambda$ can be expressed as
\begin{align*}
\frac{\dif\mathcal{T}}{\dif\lambda}=\ell+\lambda\frac{\dif\ell}{\dif\lambda}\,\,\text{  and  }\,\,\frac{\dif^2 \mathcal{T}}{\dif \lambda^2}= 2\frac{\dif\ell}{\dif\lambda}+\lambda \frac{\dif^2\ell}{\dif \lambda^2}
\end{align*}
in which $\frac{\dif\ell}{\dif\lambda}<0$ and $\frac{\dif^2\ell}{\dif \lambda^2}>0$. If $\Pi_{\lambda}$ is not empty,  $\mathcal{T}(\lambda,\theta)$ is concave for any $\lambda\in\Pi_{\lambda}$ due to $\frac{\dif^2\mathcal{T}}{\dif\lambda^2}<0$. According to the Bolzano Weierstrass theorem,  there exists a unique optimal $\overline{\lambda}_*\in\Pi_{\lambda}$ such that $\mathcal{T}(\overline{\lambda}_*,\theta)$ is maximal over $\Pi_{\lambda}$. For any  $\lambda\notin\Pi_{\lambda}$, we know $\frac{\dif^2\mathcal{T}}{\dif\lambda^2}\geq 0$ and $\frac{\dif\mathcal{T}}{\dif\lambda}\leq 0$ so that $\mathcal{T}(\lambda,\theta)$ is convex and monotonic decreasing and it does not have a maximum. Thus, $\overline{\lambda}_*$ is the sole maximizer of $\mathcal{T}(\lambda,\theta)$.
\end{IEEEproof}

\begin{figure}[!t]
	\centering
	\includegraphics[width=3.5in, height=2.25in]{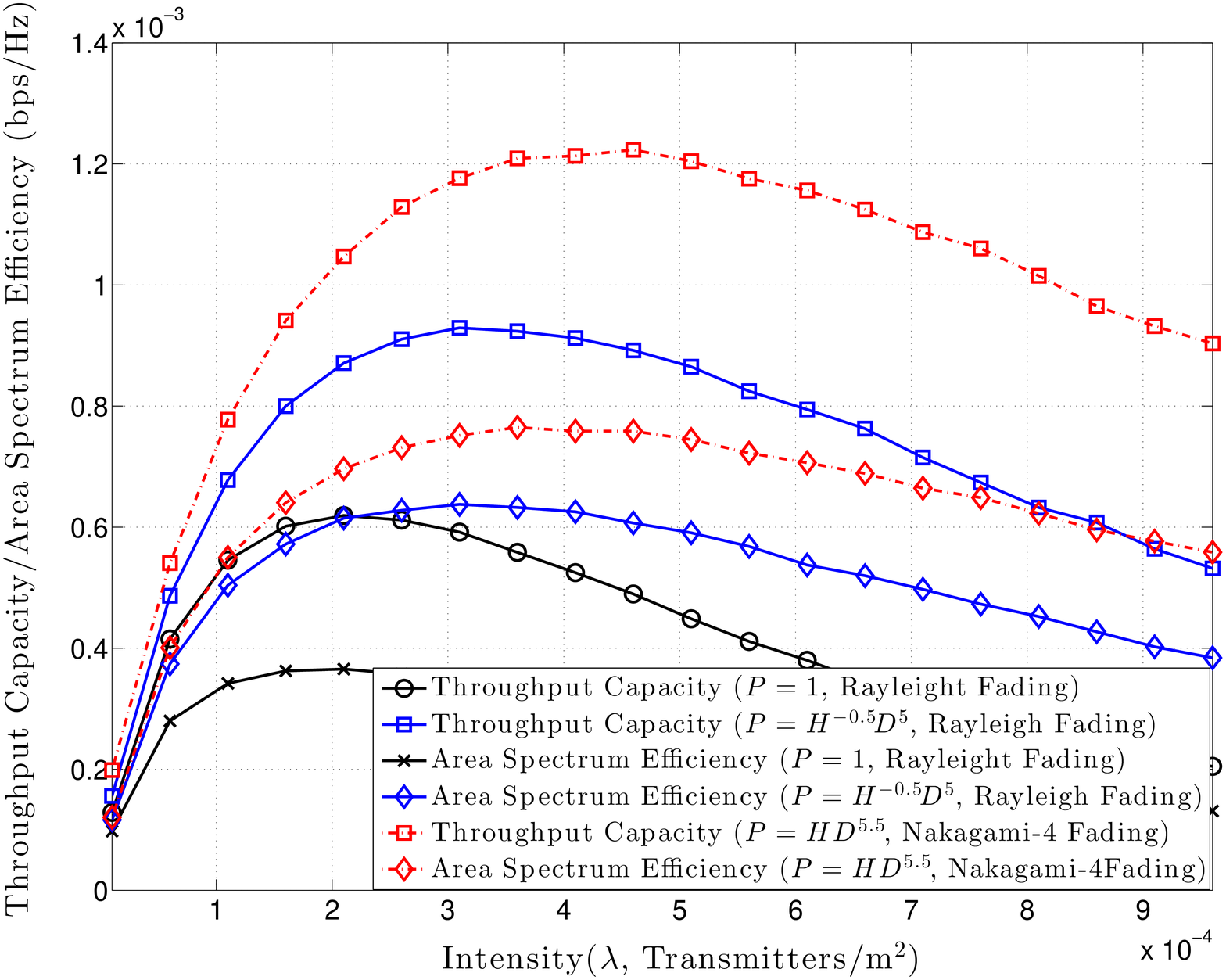}
	\caption{The simulation results of the throughput capacity ($\mathcal{T}$) and area spectrum efficiency ($\lambda C(\lambda)F^c_{\gamma_0}(\lambda,\theta)$) for $P=H^{\rho}D^{\upsilon}/\mathbb{E}[H^{\rho}D^{\upsilon}]$, $\alpha=4$, $\theta=0.5$, Nakagami-$m$ fading and uniformly distributed $D\in[15,25]$.}
	\label{Fig:MeanSIRThrputCap}
\end{figure}

In general, the optimal intensity $\overline{\lambda}_*$ cannot be found in closed-form. Nevertheless, it indeed exists in almost all of random channel gain and transmission distance models. As the simulation results shown in Fig. \ref{Fig:MeanSIRThrputCap},  there indeed exists only one optimal intensity that maximizes the throughput capacities of all power control laws. In addition, simulation results also indicate $\overline{\lambda}_*\approx \arg\sup_{\lambda}\lambda C(\lambda)F_{\gamma_0}(\lambda,\theta)$ even when $\mathcal{T}(\lambda,
\theta)$ is not a very tight bound on $\lambda C(\lambda)F_{\gamma_0}(\lambda,\theta)$.

\section{Conclusion}
The closed-form expression of the mean SIR for a Poisson  ad hoc network is found under general channel gain, transmission distance and power control models. The mean expression of SIR contains two key implications: channel randomness can benefit the mean SIR, and power control and scheduling schemes that capture the channel randomness can improve the mean SIR and outage probability.  The throughput capacity that is defined based on the mean SIR is proposed to characterize the mean area spectrum efficiency in a Shannon-capacity sense and its maximum exists if the supporting set of the unique optimal intensity of transmitters is nonempty.

\bibliographystyle{ieeetran}
\bibliography{IEEEabrv,Ref_MeanSINRAdHocNets}

\end{document}